\documentclass[12pt]{article}
\usepackage{amsfonts}
\usepackage{natbib}
\bibpunct{(}{)}{;}{a}{,}{,} %do natbib
\begin{document}

% *****************************************************************
% oznaczenia zbiorow liczbowych
% *****************************************************************

\newcommand{\rea}{{\mathbb{R}}}
\newcommand{\nnn}{{\mathbb{N}}}
\newcommand{\comp}{{\mathbb{C}}}
\newcommand{\calk}{{\mathbb{Z}}}

% *****************************************************************

% *****************************************************************
%   ogolne komendy d-spacowe
% *****************************************************************

\newcommand{\Top}[1]{\mbox{${\rm top}(#1)$}}
\newcommand{\Sc}[1]{\mbox{${\rm sc}(#1)$}}
\newcommand{\Gen}[1]{\mbox{${\rm Gen}(#1)$}}
\newcommand{\Lc}[2]{\mbox{${\rm lc}_{#1}(#2)$}}

\newcommand{\adres}[1]{  {\small\it  \begin{center} #1 \end{center}   }     }
\newcommand{\dzieki}{ \begin{center}{\bf Acknowledgments} \end{center}  }
\newcommand{\dow}{ \noindent {\bf Proof. }  }
\newcommand{\ckow}[4]{\mbox{$#1_{#2},#1_{#3},\dots ,#1_{#4}$}}
\newcommand{\tildeckow}[4]{\mbox{$\tilde{#1}_{#2},\tilde{#1}_{#3},\dots ,\tilde{#1}_{#4}$}}
\newcommand{\barckow}[4]{\mbox{$\bar{#1}_{#2},\bar{#1}_{#3},\dots ,\bar{#1}_{#4}$}}
\newcommand{\ckon}[4]{\mbox{$#1^{#2},#1^{#3},\dots ,#1^{#4}$}}
\newcommand{\tildeckon}[4]{\mbox{$\tilde{#1}^{#2},\tilde{#1}^{#3},\dots ,\tilde{#1}^{#4}$}}
\newcommand{\barckon}[4]{\mbox{$\bar{#1}^{#2},\bar{#1}^{#3},\dots ,\bar{#1}^{#4}$}}
\newcommand{\ckk}[7]{\mbox{$#1_{#2}^{#3},#1_{#4}^{#5},\dots ,#1_{#6}^{#7}$}}
\newcommand{\tildeckk}[7]{\mbox{$\tilde{#1}_{#2}^{#3},\tilde{#1}_{#4}^{#5},\dots ,\tilde{#1}_{#6}^{#7}$}}
\newcommand{\barckk}[7]{\mbox{$\bar{#1}_{#2}^{#3},\bar{#1}_{#4}^{#5},\dots ,\bar{#1}_{#6}^{#7}$}}
\newcommand{\ciag}[4]{\mbox{$#1=#2,#3,\dots,#4 $}}
\newcommand{\dsp}[1]{\mbox{$(#1,{\cal #1})$}}
\newcommand{\tildedsp}[1]{\mbox{$(\tilde{#1},\tilde{{\cal #1}})$}}
\newcommand{\bardsp}[1]{\mbox{$(\bar{#1},\bar{{\cal #1}})$}}
\newcommand{\dspa}[2]{\mbox{$(#1,{\cal #2})$}}
\newcommand{\fun}[3]{\mbox{$#1\colon #2 \- \rightarrow #3$}}

\newcommand{\genb}{\mbox{\ckow{\alpha^\bullet}{0}{1}{4}}}
\newcommand{\geno}{\mbox{\ckow{\alpha^\circ}{0}{1}{4}}}
\newcommand{\poo}{\mbox{$\cal {P}^\circ$}}
\newcommand{\zpo}{\mbox{$ {P}^\circ$}}
\newcommand{\zpto}{\mbox{$ \tilde{P}^\circ$}}
\newcommand{\pto}{\mbox{$ \tilde{\cal P}^\circ$}}
\newcommand{\pbb}{\mbox{$\cal {P}^\bullet$}}
\newcommand{\zpb}{\mbox{$ {P}^\bullet$}}
\newcommand{\zptb}{\mbox{$ \tilde{P}^\bullet$}}
\newcommand{\ptb}{\mbox{$ \tilde{\cal P}^\bullet$}}
\newcommand{\pp}{\dsp{P}}
\newcommand{\tpp}{\tildedsp{P}}
\newcommand{\ppo}{\dsp{P^\circ}}
\newcommand{\tppo}{\tildedsp{P^\circ}}
\newcommand{\ppb}{\dsp{P^\bullet} }
\newcommand{\tppb}{\tildedsp{P^\bullet}}
\newcommand{\psio}{\mbox{${\Psi}^{\circ}_{\epsilon,l,\beta}$}}
\newcommand{\psib}{\mbox{${\Psi}^{\bullet}_{\epsilon,l,\beta}$}}
\newcommand{\psitb}{\mbox{$\tilde{\Psi}^{\bullet}_{\epsilon,l,\beta}$}}
\newcommand{\psito}{\mbox{$\tilde{\Psi}^{\circ}_{\epsilon,l,\beta}$}}
\newcommand{\betab}{\mbox{${\beta}^{\bullet}$}}
\newcommand{\phb}{\mbox{$ \hat{\cal P}^{\bullet}$}}
\newcommand{\hatppb}{\mbox{$\dspa{\zpb}{\phb}$}}
\newcommand{\cbdo}{\mbox{$\Box$}}
\newcommand{\czero}{\mbox{${\cal C}_0$}}
\newcommand{\tao}{\mbox{$\tilde{A}^{\circ}$}}
\newcommand{\aoo}{\mbox{$A^{\circ}$}}
\newcommand{\tab}{\mbox{$\tilde{A}^{\bullet}$}}
\newcommand{\ab}{\mbox{$A^{\bullet}$}}
\newcommand{\boldtao}{\mbox{$\tilde{\bf A}^{\circ}$}}
\newcommand{\boldao}{\mbox{${\bf A}^{\circ}$}}
\newcommand{\boldtab}{\mbox{$\tilde{\bf A}^{\bullet}$}}
\newcommand{\boldab}{\mbox{${\bf A}^{\bullet}$}}
\newcommand{\piro}{\mbox{${\pi}_{\rho _{_{H}}}$}}
\newcommand{\tpiros}{\mbox{$\tilde{\pi}_{\rho _{_{H}}}^{\ast}$}}
\newcommand{\piros}{\mbox{${\pi}_{\rho _{_{H}}}^{\ast}$}}
\newcommand{\piroh}{\mbox{${\pi}_{\rho _{_{H}}}^{\#}$}}

% *****************************************************************
%    SRODOWISKA
% *****************************************************************

\newtheorem{tw}{Theorem}[section]
\newtheorem{lem}{Lemma}[section]
\newtheorem{stw}{Proposition}[section]
\newtheorem{defin}{Definition}[section]
\newtheorem{wn}{Corollary}[section]
% *****************************************************************
\pagestyle{myheadings}

\markright{Jacek Gruszczak: Discrete spectrum of the Deficit Angle... }

\title{\bf Discrete Spectrum of the Deficit Angle and the Differential Structure
       of a Cosmic String}
\author{Jacek Gruszczak \thanks{email: sfgruszc@cyf-kr.edu.pl} \\
Institute of Physics, Pedagogical University,\\ ul. Podchorazych 2,
       30-084 Cracow, Poland}

\date{April 18, 2006}
\maketitle
\begin{abstract}
Differential properties of  Klein-Gordon and  electromagnetic
fields on the space-time of a straight cosmic string   are
studied with the help  of methods of the differential space
theory. It is shown that these fields are smooth  in the interior
of the cosmic string space-time and that they loose this property
at the singular boundary except for the cosmic string space-times
with the following
deficit angles : $\Delta=2\pi (1-1/n) $, $n=1,2,\dots$.\\
A connection between smoothness of fields at the conical
singularity and the scalar and electromagnetic conical
bremsstrahlung is discussed. It is also argued that the
smoothness assumption of  fields at the singularity is equivalent
to the Aliev and Gal'tsov "quantization" condition leading to the
above mentioned discrete spectrum of the deficit angle.
\end{abstract}


\newpage

%Keywords: differential structure, cosmic string, quantum gravity, singularities

%********************************************************************
%                         ABSTRAKT
%^^^^^^^^^^^^^^^^^^^^^^^^^^^^^^^^^^^^^^^^^^^^^^^^^^^^^^^^^^^^^^^^^^^^

%**********************************************************************

                         \section{Introduction}

                             \label{intro}
%^^^^^^^^^^^^^^^^^^^^^^^^^^^^^^^^^^^^^^^^^^^^^^^^^^^^^^^^^^^^^^^^^^^^^
The differential structure \mbox{$\cal C  $} of a differential space (d-space
for short) \dspa{M}{C} \ is a set of real functions on $M$ which is closed with
respect to localization and closed with respect to superpositions with smooth
functions on ${\rea}^n, n\in \nnn$. Every function from ${\cal C}$ is smooth by
definition . These functions are at the base of main notions and structures
defined on \dspa{M}{C}. For instance, the smooth functions determine topology,
tangent vectors, smooth vector fields, dimension of tangent spaces, etc.
Details  can be found in \cite[]{12af}.

When one tries to describe given space-time by means of notions from the
d-spaces theory the question arises. What is the meaning of smoothness  and
d-structure  in physics of space-time? In \cite[]{7bo,12af,13af} smooth
functions from \mbox{$\cal C $} are interpreted as "the system of scalar fields
which actually contain all information necessary to define the manifold
structure". Then the first axiom of the d-space definition, postulating  the
closure of \mbox{$\cal C $} with respect to localization, guarantees the
consistency of local physics with global physics. The second axiom, postulating
the closure of \mbox{$\cal C $} with respect to superposition with smooth
functions on $\rea^{n}$, provides  a mechanism for construction "new" smooth
quantities from "old" ones. The additional, third axiom, postulating a local
diffeomorphism to $\rea^{n}$, which changes a given d-space into a smooth
manifold, can be interpreted as a "non-metric version of the equivalence
principle"  \cite[]{13af}. In the present paper, we test how this interpretation
works in practice.

The assumptions of the present paper and the main anticipated results are the
following:

1) It is supposed that  space-time under investigation is a flat space-time with
the conical singularity  usually called the space-time of straight cosmic
string.

2) Additionally, it is assumed that a scalar Klein-Gordon or electromagnetic fields
are defined on the background of this space-time, and  perturbations of the metric
due to these fields are not taken into account.

3) The detailed analysis of differential properties of the elementary solutions for
K-G scalar and  electromagnetic fields (sections \ref{dproperties},
 \ref{global} and \ref{electro})
leads to the following results: a) The elementary solutions are smooth
functions (in the sense of Sikorski) on the space-time manifold  treated as a
d-space. b) The above fields on this d-space with a conic singularity are
smooth only for the deficit angle  $\Delta = 2\pi (1-1/n),n=1,2,\dots $.

In  section \ref{A-B} the co-called scalar and electromagnetic conical
bremsstrahlung effect \cite[]{56af} is discussed. Section \ref{conclusion}
contains summary of results and an argumentation that the assumption of
smoothness of physical fields at the singularity (the asymptotic smoothness)
can be treated as a geometric version of the Aliev and Gal'tsov condition  for
vanishing of the conical bremsstrahlung effect. The assumption leads to the
following discrete spectrum of the deficit of angle: $\Delta = 2\pi
(1-1/n),n=1,2,\dots $.   In Appendix \ref{appds} one can find elementary
introduction to the theory of d-spaces, and in Appendix \ref{appcosmic} details
concerning the d-space of a cosmic string.

%********************************************************************

     \section{Differential space of a cosmic string with singularity}

                           \label{dspcosmic}

%^^^^^^^^^^^^^^^^^^^^^^^^^^^^^^^^^^^^^^^^^^^^^^^^^^^^^^^^^^^^^^^^^^^^
 Space-time described with the help of the metric
\begin{equation}
g=-dt^2+k^{-2}d\rho^2+{\rho}^2 d\phi^2 +dz^2        \label{metric}
\end{equation}
where $k=(1-\Delta/2\pi)$ , $t, z \in \rea$, $\rho\in (0, \infty)$ and $\phi\in
\langle 0, 2\pi ) $, is an example of space-time with quasiregular singularity
of the conic type. The parameter $\Delta\in \langle 0, 2\pi)$ is called {\em
deficit angle}. The three-dimensional version of the above metric is
interpreted as the Schwarzchild solution in the framework of 3-D gravity
\cite[]{37af}, whereas the four-dimensional metric is interpreted as an
exterior gravitational field of a straight cosmic string in 4-D gravity
\cite[]{41af}.

The space-time of a cosmic string as a pseudoriemannian manifold $(M, g)$ is
isometric to $(C^{\circ}\times \rea^2, \iota^*\eta^{(5)})$, where $C^\circ$ is
a two-dimensional cone without the vertex; $\iota : C^{\circ}\times \rea^2$ $
\rightarrow\rea^5$ is an  embedding and ${\eta}^{(5)}$  is the five-dimensional
Minkowski metric.  The space-time of a cosmic string
 as a differential  space  is diffeomorphic to
$C^{\circ}\times \rea^2$, where the latter is
 treated as a differential subspace of the d-space
$(\rea^5, {\cal E}_5 )$,where ${\cal E}_5 =C^{\infty} (\rea^5)$. In other
words, $(M,g)$ as a d-space \dsp{M} is diffeomorphic to $(C^{\circ}\times
\rea^2, ({\cal E}_5)_{C^{\circ}\times \rea^2})$, where $({\cal
E}_5)_{C^{\circ}\times \rea^2}$ is the induced d-structure \cite[]{16af,12af}
and the symbol  $(\cdot)_{C^{\circ}\times \rea^2}$ denotes the operation of
taking closure with respect to localization (definition \ref{loc}).

The singular space-time of a cosmic string can be defined in various ways. The
most popular method depends on attaching  the vertex of the cone to $C^{\circ}$.
Then $C^{\bullet}\times \rea^2$ represents the space-time of the cosmic string
with singularity, where $C^{\bullet}$ denotes the cone with the vertex.
Evidently, $C^{\bullet}\times \rea^2$ is not a sub-manifold of $\rea^{5}$  but
it is still a d-subspace: $(C^{\bullet}\times \rea^2, ({\cal
E}_5)_{C^{\bullet}\times \rea^2})$.

However, the main objects of our study in the present paper are two auxiliary
d-spaces \dsp{P^\circ} \ and \ \dsp{P^\bullet} \ (for details see Appendix
\ref{appcosmic} or \cite[]{16af}) which are diffeomorphic to the d-space of
cosmic string without singularity and to the d-space of cosmic string with
singularity, respectively. The auxiliary d-spaces are more convenient for
investigations than the original ones $(C^{\circ}\times \rea^2, ({\cal
E}_5)_{C^{\circ}\times \rea^2})$ and $(C^{\bullet}\times \rea^2, ({\cal
E}_5)_{C^{\bullet}\times \rea^2})$.
% The main reason is that these d-spaces represent
%also additional properties coming from the embedding.
For example, the above described singularity attaching process is simply the
procedure of taking limits of a few functions from ${\cal P}^\circ$ \
\cite[]{16af}. The resulting d-space \dsp{P^\bullet} is, in a sense, a limit
(or an asymptotic) state of the background d-space \dsp{P^\circ}.

%***********************************************************************

  \section{Differential properties of a scalar field in
                   the conical space-time of a cosmic string }

                          \label{dproperties}

%^^^^^^^^^^^^^^^^^^^^^^^^^^^^^^^^^^^^^^^^^^^^^^^^^^^^^^^^^^^^^^^^^^^^^^^

As well known,  normal modes of the Klein-Gordon scalar field for the cosmic
string space-time in $(t,\rho,\phi,z)$ coordinates have the following form

$$\fun{\tilde{\Psi}^{\circ}_{\epsilon,l,\beta}}{\tilde{P}^{\circ}}{\comp},$$
\begin{equation}\label{psito}
\tilde{\Psi}^{\circ}_{\epsilon,l,\beta}(t,\rho,\phi,z)=
N_{\epsilon,l,\beta}e^{-i\epsilon t}e^{i\beta z}e^{il\phi}\rho^{|l|/k}F(l,k;
\rho)
\end{equation}
where $\epsilon,\beta\in \rea$, $l\in\calk$, $N_{\epsilon,l,\beta}$ is the
normalization constant,  $k\in (0, 1) $ is defined in formula (\ref{metric}) and
 $\tilde{P}^{\circ}$ is given in Appendix \ref{appcosmic}. $F(l,k; \rho)  $
 is an analytical function of $\rho$. Its detailed form is not
important for the present study.

Additionally, there is an another set of normal modes usually excluded from
physical investigations because of a singular behaviour of functions as their
arguments tend to the singularity $(\rho\rightarrow 0)$ \cite[]{6bo}. Sometimes
however, such normal modes can be of physical relevance \cite[]{68af} but in the
present paper this divergent normal modes are not taken into account.

The space-time of the cosmic string is represented by the d-space \dsp{P^\circ}.
Therefore, in this case normal modes of the scalar field are given by:

$$\fun{{\Psi}^{\circ}_{\epsilon,l,\beta}}{{P}^{\circ}}{\comp},$$
\begin{equation}
\label{psio}
 {\Psi}^{\circ}_{\epsilon,l,\beta}([p]):=
\tilde{\Psi}^{\circ}_{\epsilon,l,\beta}(p),
\end{equation}
where $ p=(t,\rho,\phi,z),  \ [p]\in P^{\circ}=\tilde{P}^{\circ}/{\rho}_{_{H}}$ (
Appendix \ref{appcosmic}).

\begin{defin}
Let $(M, {\cal C})  $ be a d-space. A complex function
\fun{f}{M}\comp \
is said to be smooth if
\ ${\rm Re }f$, ${\rm Im  }f\in {\cal C }$.
\end{defin}

\begin{stw}
\label{smootho}
The normal modes
${\Psi}^{\circ}_{\epsilon,l,\beta}$
are smooth functions on \dsp{P^\circ}
 for every
$\epsilon, \beta\in\rea$,  $l\in\calk$ and $k\in (0, 1)$.
\end{stw}

\dow Following corollary \ref{smoothsfapp} it is enough to prove the smoothness
of $\tilde{\Psi}^{\circ}_{\epsilon,l,\beta}$ on \tildedsp{P^\circ} (Appendix
\ref{appcosmic}). It is easy to check that functions
$\tilde{\Psi}^{\circ}_{\epsilon,l,\beta}$ are smooth functions since they are
superpositions of generators \tildeckow{\alpha}{0}{1}{4} \ (see Appendix
\ref{appcosmic}) with smooth functions from $ C^{\infty}(\rea^m), \ m=1, 2,
\dots $ and $\rho^{|l|/k}$ is smooth for  $\rho >0$.
 \cbdo

According to  proposition \ref{smootho} the normal modes (\ref{psio}) are smooth
functions on \ppo. They belong to \poo  \ and therefore they carry no new information
from the point of view of the "pre-geometry" determined by \poo. One can say that the
space-time of a cosmic string is "prepared" for imposing a scalar field on its
background d-space. In other words, space-time is from the very beginning
differentially configured in such a manner that the imposition of a scalar field is
done by the indication which functions among already existing ones in \poo \ are
normal modes. This fact is a confirmation of the correctness of the assumption that
the scalar field on a given space-time  does not cause any changes of properties of
the gravitational field.

\ppo \ represents the background d-space of a cosmic string if \ppb is its
asymptotic state. This statement mirrors the fact that the gravitational field
described by means of  metric (\ref{metric}) is a space-time of a cosmic string
only if a singular boundary of a conic type is present.  Therefore, in the
process of calculating  normal modes, its asymptotic properties at the
singularity have to be taken into account. This kind of analysis excludes
divergent modes from further field-theory considerations \cite[]{6bo}.
 The question arises: What are differential properties of the normal
 modes at the singularity? The following argumentation clarifies the
 situation.

One can easily check that normal modes naturally prolonged to singularity,
\begin{equation}
\label{psitb} \psitb (p):=\lim_{q\rightarrow p}\psito (q), \ q\in\zpto, p\in\zptb,
\end{equation}
are constant functions on every equivalence class $[p]$  for $p\in\zptb$ (see
Appendix \ref{appds}, formula (\ref{equivalencerel})  and Appendix
\ref{appcosmic}). Therefore, they can be used for construction prolonged modes
\psib \ defined on \ppb:
\begin{equation}
\label{psib} \psib([p]):=\psitb(p), \ [p]\in \zpb , p\in\zptb .
\end{equation}
The prolongation is natural from the physical point of view.

\begin{stw}
\label{smoothb} For every $\epsilon ,\beta \in \rea$ and $l\in \calk$,
${\Psi}^{\bullet}_{\epsilon,l,\beta}$ are
\begin{itemize}
\raggedright
\item[{\bf a)}]
 smooth functions on \dsp{P^\bullet} for $k\in (0, 1)$
 such that  $| l|/k\in \nnn$,
\item[{\bf b)}]  non-smooth functions on \dsp{P^\bullet} for
$k\in (0, 1)$
 such that  $| l|/k\not\in \nnn$,
\end{itemize}
\end{stw}

\dow
It is enough to check smoothness for
${\tilde{\Psi}}^{\bullet}_{\epsilon,l,\beta}$. Functions
$e^{-i\epsilon t},$ $ e^{i\beta z},$ $ e^{il\phi},$ $ F(l,k; \rho)$
are smooth owing to the same arguments as in proposition \ref{smootho}.
The function $\rho^{| l| /k}=(\alpha_4 (\rho))^{| l| /k}$  is
a smooth function for $\rho >0$. At $\rho =0$ the superposition is
smooth the only in the case $| l| /k \in \nnn $.
\cbdo

In general,  \psib \ are not smooth functions on \ppb  except for cases when the
metric parameter $k\in (0,1)$  satisfies the condition: $ \forall \ {l}\in\calk:
|{l}|/k\in\nnn$. This means that $k=1/n$, where $n=2,3,\ldots$. One can also include
the case of the Minkowski space  ($k=1$). In other words, the prolonged normal modes
are smooth functions  only for the space-time of a cosmic string  with the deficit
angle $\Delta=2\pi (1-1/n)$, where $n=1,2,\dots$.

%*************************************************************************

\section{Global properties of the space-time of a cosmic string with a scalar field }

                             \label{global}

%^^^^^^^^^^^^^^^^^^^^^^^^^^^^^^^^^^^^^^^^^^^^^^^^^^^^^^^^^^^^^^^^^^^^^^^^^

Let a real-valued function $\betab :\zpb \rightarrow \rea$ be defined by the
formula

\begin{equation}
\label{betab}
\betab ([p]):=\tilde{\beta}^{\bullet} (p):={\rho}^{1/k},
\end{equation}
\noindent
 where
$k\in (0,1\rangle \ $. One can define an another d-structure on \zpb :
$$\hat{\cal P}^{\bullet}= {\rm
Gen}(\ckow{\alpha^\bullet}{0}{1}{4},\betab).$$

\begin{stw}
\label{global1}
For $k\neq 1/n $, $n=1,2,3,\dots$,
the prolonged normal modes \psib \ are smooth functions on \hatppb  \ for
every $\epsilon,\beta\in\rea$ and $l\in \calk$.
\end{stw}

\dow The  factor ${\rho}^{|{l}|/k}$ in formula (\ref{psito}), after
prolongation to the singular boundary, is a smooth composition of \betab . \cbdo

\begin{wn}
\label{global2} If $k=1/n $,  $n=1,2,3,\dots$ then \hatppb=\ppb.
\end{wn}
\dow $\betab\in\pbb$ for $k=1/n$. \cbdo

\begin{stw}
\label{global3} If $k\neq 1/n$,  $n=1,2,3,\dots$ then the set $\{
\ckow{\alpha^\bullet}{0}{1}{4},\betab\} $ is differentially independent at
boundary points $p\in{\bf S}=\zpb-\zpo$ . \ \betab \ and
${\alpha}^{\bullet}_{4}$ differentially depend on
$\{{\alpha}^{\bullet}_{0},{\alpha}^{\bullet}_{1},{\alpha}^{\bullet}_{2},
{\alpha}^{\bullet}_{3}  \}$ elsewhere on \zpb .
\end{stw}
\dow The conclusion is a straightforward consequence of  definitions
\ref{depen} and \ref{inde}. \cbdo

\begin{wn}
\label{global4}
If $k\neq1/n$ then
\begin{itemize}
\raggedright
\item[{\rm a)}] ${\rm dim}\ T_{p}\hatppb =6$ for $p\in {\bf S}$,

\item[{\rm b)}]${\rm dim}\ T_{p}\hatppb =4$ for $p\not\in {\bf S}$,
\end{itemize}
where ${\bf S}=\zpb-\zpo$ \ denotes the set of all boundary points.
\end{wn}
\dow The conclusion is a consequence of lemma \ref{wym} and proposition
\ref{global3} \cbdo \\

In general, \ \psib \ are non-smooth functions on \ppb because of the \
${\rho}^{|l|/k}$ factor in   formula (\ref{psito}). Other factors are smooth
functions. In order to keep smoothness of \psib \ on \zpb \ one has to modify
the d-structure \pbb \ by adding "necessary" functions. In our case, the most
physically reasonable method is to supplement the set of generators \{\genb  \}
\ of the d-structure \pbb \ (Appendix \ref{appcosmic}) with the function \betab
. Then $\hat{\cal P}^{\bullet}= {\rm
Gen}(\ckow{\alpha^\bullet}{0}{1}{4},\betab)$ is the smallest d-structure
containing \genb \ and \betab \ as  smooth functions \cite[]{12af,16af}.

Thus, in order to keep smoothness of \psib \ , the prolonged d-space of the
cosmic string space-time  with a scalar field ought to be represented by \hatppb
\ rather than by \ppb . In the case $k=1/n$, the function $\beta^\bullet$  is
smooth on \ppb \ and according to corollary \ref{global2} the prolonged
background d-spaces both for the cosmic string space-time and the cosmic string
space-time with a scalar field are the same; $\hatppb = \ppb$. In the case
$k\not =1/n$,  $\beta^\bullet$ is not a smooth function on \ppb. This means
that \hatppb \ and \ppb \ are different and are not diffeomorphic d-spaces. For
example \hatppb \ has differential dimension 6 at the singular points (corollary
\ref{global4}) whereas the \ppb ---  5 (see \cite[]{16af}). A straightforward
consequence of this fact is that \hatppb \ cannot be embedded in $\rea^5$ like
\ppb, but can be embedded in $\rea^6$.

In order to formulate the final conclusions of the above discussion let
us define the following.

\begin{defin}
\label{smoothphf}
A physical scalar (vector, tensor) field $\Phi$ is said to be smooth on
a background d-space \dspa{M}{C} \ iff scalar (vector, tensor)
elementary solutions of the corresponding field equation are smooth
functions (vector, tensor fields) on \dspa{M}{C}.
\end{defin}

\begin{defin}
\label{asmoothphf} A physical scalar (vector, tensor) field $\Phi$ is said to be
 asymptotically smooth on $(M^{\circ}, {\cal C}^{\circ} ,M^{\bullet}, {\cal
C}^{\bullet} , g )$ if
\begin{itemize}
\item[{\rm a)}] $\Phi$ is a smooth physical field on
$(M^{\circ}, {\cal C}^{\circ})$ ,
\item[{\rm b)}] $\Phi$ is a smooth physical field on
$(M^{\bullet}, {\cal C}^{\bullet})$ ,
\end{itemize}
where the symbol $(M^{\circ}, {\cal C}^{\circ} ,M^{\bullet}, {\cal C}^{\bullet}
, g )$ denotes a pseudoriemannian manifold $(M^{\circ} , g )$ equipped with a
metric $g$ which, as a d-space $(M^{\circ}, {\cal C}^{\circ})$,  has the
prolongation $(M^{\bullet}, {\cal C}^{\bullet})$.
\end{defin}

The prolonged background d-space of a cosmic string is represented strictly by
\ppb \ (see Appendix \ref{appcosmic}) so  \hatppb \ cannot be interpreted in
such manner. Thus, including Minkowski space-time ($k=1$), one can formulate the
following theorem.

\begin{tw}
\label{nostring} The conical space-time of a cosmic string $(\zpo ,\poo ,\zpb
,\pbb , g )$ can be a background of an asymptotically smooth Klein-Gordon
scalar field   only in the case of the following discrete spectrum of the
deficit  angle:
$$\Delta=2\pi (1-1/n), $$
where $n=1,2,\dots  $, and $g$ denotes  metric {\rm (\ref{metric})}.
\end{tw}

I would like to emphasize that the discrete spectrum of the deficit  angle is
not visible on the "metric" level. The effect appears when the differential
properties of a cosmic string are taken into account.

%************************************************************************
                  \section{Electromagnetic field in a conical space-time}

                           \label{electro}
%^^^^^^^^^^^^^^^^^^^^^^^^^^^^^^^^^^^^^^^^^^^^^^^^^^^^^^^^^^^^^^^^^^^^^^^^

Let  \fun{\tao_{\sigma}}{\zpto }{\comp } be an electromagnetic field.
$\tao_{\sigma}$, in the Lorentzian gauge   $\nabla^{\mu}\tao_{\mu}=0$, obey the
Maxwell equations $\nabla_{\mu}\nabla^{\mu} \tao_\sigma =0$. Their elementary
solutions have the following form

\begin{itemize}
\item[]$ \tao_{a} (\epsilon,\beta, l )= e^{-i\epsilon t}e^{i\beta z}e^{il\phi}\rho^{|
l| /k} { F}^{a}_{{\epsilon ,\beta , l,k}}(\rho ),$

\item[]$  \tao_{1}
(\epsilon,\beta, l )= e^{-i\epsilon t}e^{i\beta z}e^{il\phi} \left(\rho^{|1+ l
/k|} { F}^{1}_{{\epsilon ,\beta , l,k}}(\rho )+\rho^{|1- l /k|} {
F}^{2}_{{\epsilon ,\beta , l,k}}(\rho )\right),$

\item[] $\tao_{2}
(\epsilon,\beta, l)= \frac{k\rho}{2i} e^{-i\epsilon t}e^{i\beta z}e^{il\phi}
\left(\rho^{|1+ l /k|} { F}^{1}_{{\epsilon ,\beta , l,k}}(\rho )-\rho^{|1- l
/k|} { F}^{2}_{{\epsilon ,\beta , l,k}}(\rho )\right),$
\end{itemize}

where   $\epsilon , \beta \in\rea $,  $l\in\calk $ and a=0, 3. Detailed forms
of the analytical functions ${ F}^{b}_{{\epsilon ,\beta , l,k}}$, $b=0,1,2,3$
can be found in \cite[]{56af} but  they are not relevant  for further
discussion.

\begin{defin}

Let ${\cal B}_{1}$  and \ ${\cal B}_{2}$ be two sets of real functions on $M$ \ and
$\fun{{\bf V}_{m}}{{\cal C}}{{\cal B}_{m}}$, ${\cal B}_{m}={\bf V}_{m} ({\cal C})$,
$m=1,2$, be  vector fields on \dspa{M}{C}. The mapping ${\bf V}:={\bf V}_{1}+i{\bf
V}_{2}$ is said to be a complex vector field on  \dspa{M}{C}. ${\bf V}$ is a smooth
complex vector field on  \dspa{M}{C} if ${\cal B}_{m}\subset {\cal C}$ for $m=1,2$.
\end{defin}

Let us define complex vector fields $\boldtao_{{\epsilon ,\beta , l}}$   and
$\boldtab_{{\epsilon ,\beta , l}}$  on \tppo \ and \tppb, respectively, by the
following formulae
$$\boldtao_{{\epsilon ,\beta , l}} :={\tao}^{\mu}{\partial}_{\mu},$$
$$\boldtab_{{\epsilon ,\beta , l}}({\gamma}^\bullet)(p) :=\lim_{q\rightarrow
p}\boldtao_{{\epsilon ,\beta , l}} ({\gamma}^{\circ})(q),$$
 where
$\gamma^\circ :=\gamma^\bullet\mid_{\zpto}$, $\gamma^{\bullet}\in\ptb$,
$q\in\zpto$ and $p\in\zptb$.

Then, electromagnetic vector fields $\boldao_{{\epsilon ,\beta , l}}$   and
$\boldab_{{\epsilon ,\beta , l}}$  on \ppo \ and \ppb, respectively, can be
defined in the following way
$$\boldao_{{\epsilon ,\beta , l}} := \piroh (\boldtao_{{\epsilon ,\beta , l}}),$$
$$\boldab_{{\epsilon ,\beta , l}} :=\piroh (\boldtab_{{\epsilon ,\beta , l}}),$$
where the map   \piroh \  is given in Appendix \ref{appds}.

\begin{stw}

\label{smoothA}
For every  $\epsilon, \beta\in\rea  $ and $ l\in\calk$
\begin{enumerate}
\item  \label{smoothAo}
$\boldao_{{\epsilon ,\beta , l}}$ \ are smooth complex vector fields on \ppo,
\item \label{smoothAb}
$ \boldab_{{\epsilon ,\beta , l}}$ \ are
\begin{itemize}
\raggedright
\item[{\rm a)}]
 smooth complex vector fields on \dsp{P^\bullet}, for $k\in (0, 1)$,
 such that  $| l|/k\in \nnn$,

\item[{\rm b)}]  not smooth complex vector fields on \dsp{P^\bullet}, for
 $k\in (0,1)$, such that  $| l|/k\not\in \nnn$.
\end{itemize}
\end{enumerate}
\end{stw}
\dow Following corollary \ref{smoothvf},
 \ $\boldao_{{\epsilon ,\beta , l}}$ and
$\boldab_{{\epsilon ,\beta , l}}$ are smooth vector fields on \ppo \ and \ppb,
respectively, iff both $\boldtao_{{\epsilon ,\beta , l}}$ on  \tppo \ and
$\boldtab_{{\epsilon ,\beta , l}}$ on \tppb \ are smooth. Straightforward
calculations  lead to the conclusion that ${\rm Re} \ \boldtao_{{\epsilon ,\beta
, l}}(\pto )\subset\pto $, ${\rm Im} \ \boldtao_{{\epsilon ,\beta , l}}(\pto
)\subset\pto $, for every $k\in (0,1)$, and ${\rm Re} \ \boldtab_{{\epsilon
,\beta , l}}(\ptb )\subset\ptb $, ${\rm Im} \ \boldtab_{{\epsilon ,\beta ,
l}}(\ptb )\subset\ptb $  only for $k\in (0,1)$ such that $| l|/k\in \nnn$ for
every $l\in \calk $. \cbdo

\begin{tw}
\label{nostringA}

The conical space-time of a cosmic string $(\zpo ,\poo ,\zpb ,\pbb , g )$ can
be a background of an asymptotically smooth electromagnetic field   only in the
case of the following discrete spectrum of the deficit  angle:
$$\Delta=2\pi (1-1/n), $$
where $n=1,2,\dots  $, and $g$ denotes  metric {\rm (\ref{metric})}.

\end{tw}

\dow The argumentation is similar to that of in the scalar field case. Following
proposition \ref{smoothA} and definition \ref{asmoothphf},
$\boldab_{{\epsilon,\beta , l}}$ are not smooth complex vector fields on \ppb
for $k\neq 1/n$, $n=1,2,\dots$ and therefore they are not asymptotically smooth
on $(\zpo ,\poo ,\zpb ,\pbb , g )$. For $k=1/n$, $n=1,2,\dots$ both
$\boldao_{{\epsilon ,\beta , l}}$ and $ \boldab_{{\epsilon ,\beta , l}}$ are
smooth complex vector fields on \ppo \ and \ppb \, respectively. \cbdo

%**********************************************************************

              \section{Radiative Aharonov-Bohm effect and differential structures}

                             \label{A-B}

%^^^^^^^^^^^^^^^^^^^^^^^^^^^^^^^^^^^^^^^^^^^^^^^^^^^^^^^^^^^^^^^^^^^^^^
The space-time of a cosmic string is locally flat and consequently there are no
local gravitational forces acting on massive bodies or light rays. In spite of
this, there are a few interesting effects such as: the lensing effect
\cite[]{10af,51af}, production of an electromagnetic radiation by a freely
moving charge \cite[]{8bo}, radiative "conical bremsstrahlung"
\cite[]{56af,43af},
  which are examples of the gravitational
Aharonov-Bohm effects \cite[]{67af,51af}. From the point of view of the present
paper the most interesting are the so-called radiative A-B effects appearing
when a scalar or charged particle is moving in the space-time of a cosmic string
\cite[]{56af}.

A scalar (electric) charge freely moving in the space-time of a cosmic string
can be regarded as a source of a scalar (electromagnetic) field with
non-vanishing energy-momentum tensor. During the motion  a variation of the
total energy of the field appears. The variation ${\cal E}$ is interpreted by
Aliev and Gal'tsov as the total work done by a radiation friction force upon the
source. The effect is called {\em scalar (electromagnetic) conical
bremsstrahlung} \cite[]{56af}.

In the scalar and electromagnetic cases, the distribution of variation of the
total energy is of the form
$$\frac{d{\cal E}}{d\omega}=\frac{{\sin}^2(\pi
/k)}{\pi k}F_{sc \ (em)}(k,q,\omega,d,v,U^0),$$ where $F_{sc \ (em)}$ is a
function of $k,\omega$ \ and constants of motion. Its detailed form can be
found in the original paper by Aliev and Gal'tsov \cite[]{56af}. However, for
our purposes only the dependence on $k$ \ is important. In both scalar and
electromagnetic cases, ${\cal E}$ vanishes for $k=1/n$ ($\Delta=2\pi (1-1/n)$),
$n=1,2,\dots$.

Thus, there is an apparent connection between the smoothness of the elementary
solutions on the d-space of a cosmic string with singularity \ppb \ and the
effect of vanishing of ${\cal E}$. The nature of this connection has a
relatively simple mathematical origin. Namely, the variation of the total
energy ${\cal E}$ is calculated by means of so-called radiative Green function
which is constructed with the help of the elementary solutions \psitb \ in the
scalar case and with the help of $\tilde{A}^\bullet_{\mu }(\epsilon , \beta ,
l)$ in the electromagnetic case. As  shown in \cite[]{43af,56af,8bo},  ${\cal
E}$ \ constructed in such a manner vanishes for smooth elementary solutions on
\ppb,
 \ and is different from zero for non-smooth ones.

%************************************************************************
                  \section{Summary and discussion}

                           \label{conclusion}
%^^^^^^^^^^^^^^^^^^^^^^^^^^^^^^^^^^^^^^^^^^^^^^^^^^^^^^^^^^^^^^^^^^^^^^^^

The main  purpose of the present paper  is to test whether, and in what way,
physical fields on the space-time of a cosmic string  participate in the
formation of the  manifold structure \poo \ and the d-structure \pbb , where
\pbb \ represent the d-structure  of the space-time of a cosmic string  with the
singular boundary (see section \ref{intro} and \cite[]{13af,63af}).

Mathematically, the test is based on verifying whether the elementary solutions
of a scalar field belong to \poo \ or, after prolongation to $\pbb$. In the case
of an electromagnetic field, one tests the smoothness of $\boldao_{{\epsilon
,\beta , l}}$   and $\boldab_{{\epsilon ,\beta , l}}$  on \ppo \ and \ppb,
respectively.

Propositions \ref{smootho} and \ref{smoothA}.\ref{smoothAo} state that, indeed,
the physical fields in the interior of the cosmic string space-time \zpo \
participate in the formation of the manifold structure \poo \ in such a way that
they can be reconstructed by means of the original space-time generators
$\{\geno \}$ by using  the operation of taking closure with respect to
superposition with smooth functions on $\rea ^n$ (see Appendix  \ref{appds}).
This is consistent  with the interpretation mentioned in section \ref{intro} or
in \cite[]{13af}.

A new situation appears when one takes into consideration the cosmic string
space-time with singularity. Such an object is not a manifold, but it  is still
a d-space \ppb. With the exception of the cosmic string space-times with
$\Delta=2\pi (1-1/n)$, $n=1,2,\dots$,  scalar and electromagnetic fields do not
participate in the formation  of the original d-structure \pbb (propositions
\ref{smoothb} and
 \ref{smoothA}.\ref{smoothAb}).

Thus, if one assumes that space-time of a cosmic string is a pseudoriemannian
manifold $(\zpo , g)$ \ which, as the d-space \ppo, \ has the prolongation \ppb
\ (section \ref{dspcosmic} and definition \ref{asmoothphf}) then it can be a
background of an asymptotically smooth scalar field or of  an asymptotically
smooth electromagnetic field only for the following deficit  angles $\Delta =
2\pi (1-1/n)$, $n=1,2,\dots$, (theorems \ref{nostring} and \ref{nostringA}).

However, it is also interesting  to test what happens if the elementary
solutions are assumed to be smooth functions on the whole of $\zpb$ even in the
case $\Delta\ne 2\pi (1-1/n)$, $n=1,2,\dots$. In the present paper the
consequences of such an assumption were discussed for the case of a scalar K-G
field (section \ref{global}). It turns out  that the assumption of smoothness of
the scalar elementary solutions (normal modes) is satisfied when the
d-structure on \zpb \ is \phb \ instead of $\pbb$.  \phb \ is the smallest
d-structure (in the sense of inclusion) which contains the scalar elementary
solutions. The space-time with singularity $\zpb$  and the d-structure \phb \
forms a d-space \hatppb \ which is not diffeomorphic to the original background
d-space with singularity $\ppb$. For example,  \hatppb \ differs from $\ppb$
 by its embedding properties. It cannot be embedded in $\rea^5$
like $\ppb$ (corollary \ref{global4}). It is worth emphasizing that both \hatppb
\ and \ppb \ have the same topology and the same metric (\ref{metric}) defined
on $\zpo\subset\zpb$.

One can wonder whether the asymptotic smoothness (smoothness at the singularity)
plays any role in the context of physical investigations. But, from the
mathematical point of view, the  asymptotic smoothness requirement for physical
fields is well motivated since the smoothness is a key notion within the theory
of d-spaces and the space-time with singularity is a d-space. Smooth objects
define a d-space's properties. One can say that non-smooth objects are
"outside" the d-space's theory. In a sense,  "non-smoothness" is a symptom of
the theory inconsistency.

If one tries to model  physical reality with the help a d-space then every
physical field has to be smooth. Therefore, the asymptotic non-smoothness of
the considered physical fields for $\Delta\ne 2\pi (1-1/n)$, $n=1,2,\dots$, is
a serious defect which is non removable without modifications of the
d-structure \pbb. Thus, the consistency assumption   of the  theory of physical
fields (scalar and electromagnetic) on the cosmic string space-time  in the
context of the d-spaces theory leads to the following deficit angle
"quantization" condition: $\Delta = 2\pi (1-1/n)$, $n=1,2,\dots$.

One can  compare the above results  with the conclusions obtained by Aliev and
Gal'tsov (section \ref{A-B} or \cite[]{56af}). The scalar and electromagnetic
conical bremsstrahlung occurs only in the case of asymptotically non smooth
scalar and electromagnetic elementary solutions on $\ppb$. In other words, the
radiative scalar or electromagnetic A-B effects vanish under the assumption of
asymptotic smoothness of  solutions. The disappearance of the conical
bremsstrahlung was treated by Aliev and Gal'tsov as a "quantization" condition
analogously to the well known effect for the quantum-mechanical A-B effect for a
magnetic flux. The asymptotic smoothness assumption plays a similar role.
%************************************************************************
%                   PODZIEKOWANIA
%^^^^^^^^^^^^^^^^^^^^^^^^^^^^^^^^^^^^^^^^^^^^^^^^^^^^^^^^^^^^^^^^^^^^^^^^
\dzieki I thank Prof. Michael Heller and Prof. Krzysztof Ruebenbauer for their
comments, valuable discussion and suggestions.
%@@@@@@@@@@@@@@@@@@@@@@@@@@@@@@@@@@@@@@@@@@@@@@@@@@@@@@@@@@@@@@@@@@@@@@
%                       APPENDIKSY
%@@@@@@@@@@@@@@@@@@@@@@@@@@@@@@@@@@@@@@@@@@@@@@@@@@@@@@@@@@@@@@@@@@@@@@

%\noindent{ \Large \bf Appendixes}
%\appendix
\newpage
\begin{appendix}

%************************************************************************

       \section{Differential spaces}

                               \label{appds}
%^^^^^^^^^^^^^^^^^^^^^^^^^^^^^^^^^^^^^^^^^^^^^^^^^^^^^^^^^^^^^^^^^^^^^^^^

The fundamental notions and theorems of the theory of differential spaces in the
sense of Sikorski  can be found in a monograph by R. Sikorski \cite[]{3bo} or in
\cite[]{12af,15af,16af,34af,35af}. Here I give the  definitions and theorems
necessary to follow the present paper. Informations about spaces more general
than d-spaces in the sense of Sikorski  can be found in
\cite[]{1af,27af,28af,36af,59af,60af,61af}.

Let \czero \ be a set of real functions on $M$.
\begin{defin}
The set of functions
$${\rm sc}(\czero):=\{f=\omega (\varphi_1,\varphi_2,\dots
,\varphi_n): \ \ \varphi_1,\varphi_2,\dots,\varphi_n\in \czero,\- \ \ \omega\in {\rm
C}^\infty (\rea^n),\- \ \ n\in\nnn \} $$ is said to be the closure with respect to
superposition with smooth functions from ${\rm C}^\infty (\rea^n)$\ for every
$n\in\nnn$.
\end{defin}

A function \fun{f}{M}{\rea} is local \czero -function if for every
$p_0\in M$ there is a neighbourhood $U\in{\rm top}(M)$  and
$\varphi\in\czero$ such that $f|_U=\varphi|_U$.

\begin{defin}
\label{loc}
The set of all local \czero -functions on $M$ denoted by
$ (\czero)_{_M} $
is called the closure with respect to localization.
\end{defin}
 Details  can be found in
\cite[]{3bo,16af}.

\begin{defin}
The set
$ {\cal C}={\rm Gen}(\czero):=({\rm sc}(\czero))_{_M} $
is said to be generated by \czero . Then the \czero\ is
called the set of generators.
\end{defin}

\begin{tw}
Let \czero \ be a set of real functions on $M$ and ${\cal C}$ the set of
functions generated by \czero ; ${\cal C}:={\rm Gen}(\czero)$. Then
\begin{itemize}
\item[] {\rm 1)} ${\rm top}({\cal C})={\rm top}(\czero)$,
\item[] {\rm 2)} ${\cal C}$ is a d-structure,
\item[] {\rm 3)} $\czero \subset {\cal C}$,
\item[] {\rm 4)} ${\cal C}$\ is the smallest (in the sense of inclusion)
d-structure containing \czero.
\end{itemize}

\end{tw}

Proof  can be found \cite[]{12af}. \cbdo

\begin{defin}
If the set of generators \czero \ of a d-structure ${\cal C}$\ is finite then the
resulting d-space \dspa{M}{{\cal C}} \ is said to be finitely generated.
\end{defin}

% terazd rel rown
Let \dspa{M}{C} \ be a d-space. One can define the following equivalence
relation
\begin{equation}
\label{equivalencerel} \forall p, q \in M : p \ \rho_{_{H}} q \Leftrightarrow \forall
\alpha\in {\cal C}:\alpha (p)=\alpha (q).
\end{equation}
If an equivalence class (with respect to $\rho _{_{H}}$) $[p]\neq \{p \}$ for  $p \in
M$,  the topological space $(M, {\rm top}({\cal C}))$ is not Hausdorff.

Let  \piro  \ and \tpiros \ denote the following maps

\begin{list}{}{\leftmargin=1.5in}
\item[]$\fun{\piro}{M}{M/\rho_{_{H}}} , \ \piro (p)=[p],$
\item[]$\fun{\tpiros}{ {\rea}^{M/\rho_{_{H}}} }{ {\rea}^{M} }, \  \tpiros (\alpha)=
\alpha\circ\piro,$
\end{list}

where $\alpha\in\rea^{M/\rho_{_{H}}}$.

The set ${\cal C}/\rho_{_{H}} :={\tpiros}^{-1}({\cal C})$ of real functions on
$M/\rho_{_{H}}$ forms a d-structure ${\cal C}/\rho_{_{H}}$ on $M/\rho_{_{H}}$. The
d-structure is said to be
   coinduced d-structure   from
${\cal C}$ by the mapping \tpiros \ \cite[]{58af}.

Thus, the quotient space $M/\rho_{_{H}}$ can be equipped with the d-structure
${\cal C}/\rho_{_{H}}$ forming a Hausdorff d-space $(M/\rho_{_{H}} , {\cal
C}/\rho_{_{H}})$. It is easy to see that $$\fun{\piro}{M}{M/\rho_{_{H}}}$$ \ is
a smooth mapping between $(M,{\cal C})$ and $(M/\rho_{_{H}} , {\cal
C}/\rho_{_{H}})$  and, in addition,
$$\piros :=\tpiros\mid _{{\cal C}/\rho_{_{H}}}, \
\fun{\piros}{{\cal C}/\rho_{_{H}}}{{\cal C}}$$ is the isomorphism of algebras
${\cal C}/\rho_{_{H}}$ and ${\cal C}$ \cite[]{30af}. This result is very
useful  in the following form
\begin{wn}
\label{smoothsfapp}
 $\Phi$ is a smooth function on $(M/\rho_{_{H}} , {\cal C}/\rho_{_{H}})$ iff
\   $\tilde{\Phi}:=\piros (\Phi) $  is a smooth function on $\dspa{M}{C}$.
\end{wn}

% koniec
Let \tildedsp{P} be a finitely generated d-space with the d-structure
$\tilde{\cal P}$ generated by the set of functions \{
\tildeckow{\alpha}{1}{2}{n} \}; $\tilde{\cal P}=Gen\{
\tildeckow{\alpha}{1}{2}{n} \}$, \fun{\tilde{\alpha}_{k}}{\tilde{P}}{\rea},
\ciag{i}{1}{2}{n}. Then $(\tilde{P}/\rho_{_{H}} , \tilde{\cal P}/\rho_{_{H}})$
is also a finitely generated d-space with d-structure $\tilde{\cal
P}/\rho_{_{H}} ={\rm Gen}(\ckow{\alpha}{1}{2}{n})$ \ where
\fun{{\alpha}_i}{\tilde{P}/\rho_{_{H}}}{\rea}, $i=1,2,\dots n$ are given by the
following formula $\piros (\alpha_{i})=\tilde{\alpha}_{i}$. In other words
$\alpha _{i}([p]):=\tilde{\alpha}_{i}(p)$, $p\in\tilde{P}$,
$[p]\in\tilde{P}/\rho_{_{H}}$.

\begin{defin}
Let ${\cal B}$ be a set of real functions on $M$.
A linear mapping \fun{\bf V}{{\cal C}}{{\cal B}} \ such that
$${\bf V}(\alpha\beta)={\bf V}(\alpha)\beta+\alpha {\bf V}(\beta),$$
for any $\alpha,\beta\in{\cal C}$, is said to be a vector field on \dspa{M}{C}. A
vector field is smooth  if ${\cal B}\subset {\cal C}$.
\end{defin}

The set of all smooth vector fields on a d-space \dspa{M}{C} \ is a module over
$\rea$  \ and is denoted by ${\bf X}(M)$.

Let \fun{\piro}{M}{M/\rho_{_{H}}} \ and \fun{\piros}{{\cal C}/\rho_{_{H}}}{{\cal C}}
\ be the mappings as above. Then the mapping
\begin{list}{}{\leftmargin=1.5in}
\item[]$ \fun{\piroh }{{\bf X}(M)}{{\bf X}(M/\rho_{_{H}})}$,
\item[]$ \piroh ({\bf V}):=\piros ^{-1}\circ {\bf V}\circ\piros , \ {\bf V}\in {\bf
X}(M)$
\end{list}
is an isomorphism of modulae ${\bf X}(M)$ and ${\bf X}(M/\rho_{_{H}})$
\cite[]{32af}. Let us formulate this result in form useful in the present paper.
\begin{wn}
\label{smoothvf}
 ${\bf \tilde{V}}$
is a smooth vector field on \dspa{M}{C} \ iff \ ${\bf V}:=\piroh ({\bf \tilde{V}})$ \
is smooth  on $(M/\rho_{_{H}},{\cal C}/\rho_{_{H}})$.
\end{wn}

\begin{defin}
\label{depen} Let \dspa{M}{C} \  be any d-space. A function \fun{\beta}{M}{\rea}  \
is said to be differentially dependent on functions $\ckow{\alpha}{1}{2}{n}\in{\cal
C}$ \ at a point $p\in M$ if there exist a neighbourhood $U\in \Top{{\cal C}}$ of $p$
and a function $\omega\in {\rm C}^{\infty}(\rea^n)$ \ such that
\begin{eqnarray}
\beta |_U = \omega (\ckow{\alpha}{1}{2}{n})|_U .
\nonumber
\end{eqnarray}
\end{defin}

\begin{defin}
\label{inde} A set $\{\ckow{\alpha}{1}{2}{n}\}\subset {\cal C}$ \ is said to be
differentially independent  (d-independent) at a point $p\in M$ \ if no function
$\alpha_i$, for $i\in \{1,2,\dots ,n\}$, \ depends differentially on the remaining
functions at $p$.
\end{defin}

\begin{lem}
\label{wym}
Let \dspa{M}{C}  \ be a d-space with the d-structure ${\cal C}$ \
generated by the set of functions $\{\ckow{\alpha}{1}{2}{n}\}$.
The set of functions $\{\ckow{\alpha}{1}{2}{n}\}$ \ is d-independent at
$p\in M$ \ iff ${\rm dim} \ T_p M=n$
\end{lem}
\noindent Proof can be found in \cite[]{55af}. See also \cite[]{29af,16af}.
\cbdo

%\def\thesection{ Appendix \Alph{section}}

%*******************************************************************
\newpage
    \section{Definitions and Formulae for the Cosmic String Space-Time}

                              \label{appcosmic}

%^^^^^^^^^^^^^^^^^^^^^^^^^^^^^^^^^^^^^^^^^^^^^^^^^^^^^^^^^^^^^^^^^^^

Let \tppo \ be an auxiliary d-space, where
$\tilde{P}^\circ:=\rea\times
(0,\infty)\times \langle 0,2\pi\rangle \times \rea$
is a "parameter space",
$\pto :={\rm Gen}(\tildeckow{{\alpha}^{\circ}}{0}{1}{4})$
 and functions
\fun{{\tilde\alpha}^{\circ}_{i}}{\tilde{P}^\circ}\rea, $i=0,1,\dots ,4$ are given by
 the following formulae
\begin{eqnarray}
{\tilde\alpha}^{\circ}_{0}(\tilde p):=t, \nonumber\\
{\tilde\alpha}^{\circ}_{1}(\tilde p):=\rho\cos\phi,  \nonumber\\
{\tilde\alpha}^{\circ}_{2}(\tilde p):=\rho\sin\phi, \nonumber\\
{\tilde\alpha}^{\circ}_{3}(\tilde p):=z, \nonumber\\
{\tilde\alpha}^{\circ}_{4}(\tilde p):=\rho, \nonumber
\end{eqnarray}
 where
 $\tilde{p}\in \tilde{P}^\circ $. \tppo \ is not Hausdorff.
 The finitely generated d-space
 \dsp{P^\circ},
$P^\circ=\tilde{P}^\circ/\rho_{ _{H}}$, $\mbox{$\cal
P$}^\circ=\tilde{\mbox{$\cal P $}}^\circ/\rho_{ _{H}}:= {\rm
Gen}(\ckow{\alpha^\circ}{0}{1}{4})$, \fun{\alpha^{\circ}_{i}}{P^\circ}{\rea},
${\alpha}^{\circ}_{i}([p]):={\tilde\alpha}^{\circ}_{i}(p) $, $i=0,1,\dots ,4$
(see Appendix \ref{appds}), is a Hausdorff topological space and the following
lemma holds

\begin{lem}
 The d-space \dsp{P^\circ} is diffeomorphic to
      $(C^{\circ}\times \rea^2,
      ({\cal E}_5)_{C^{\circ}\times \rea^2})$, where ${\cal E}_5 =C^{\infty} (\rea^5)$.
\end{lem}

\noindent Proof  can be found in \cite[]{16af}. \cbdo

A similar lemma is valid for the cosmic string space-time with singularity. Let \tppb
\ be an auxiliary prolonged d-space, where $\tilde{P}^\bullet:=\rea\times \langle
0,\infty)\times \langle 0,2\pi\rangle \times \rea$, $\ptb :={\rm
Gen}(\tildeckow{{\alpha}^{\bullet}}{0}{1}{4})$ and
\fun{\tilde{\alpha}^{\bullet}_{i}}{\tilde{P}^\bullet}{\rea} are defined as follows
$$\tilde{\alpha}^{\bullet}_{i}(\tilde{p}_\bullet):=
\lim_{\tilde{p}\rightarrow\tilde{p}_\bullet}
{\tilde\alpha}^{\circ}_{i}(\tilde{p}), $$ where
$\tilde{p}\in\tilde{P}^\circ,\tilde{p}_\bullet\in \tilde{P}^\bullet$ and $i=
0,1,\dots ,4$. \tppb  \ is also not Hausdorff.

Let
$P^\bullet=\tilde{P}^\bullet/\rho_{ _{H}}$
and
$\mbox{$\cal P$}^\bullet
=\tilde{\mbox{$\cal P  $}}^\bullet/\rho_{ _{H}}
={\rm Gen}(\ckow{\alpha^{\bullet}}{0}{1}{4})$,
 $\alpha^{\bullet}_{i}([p])=\tilde{\alpha}^{\bullet}_{i}$,
 $i=$ $0,1,...,4$. The differential space \dsp{P^\bullet} \ is a Hausdorff
 topological space.
\begin{lem}
The d-space
\dsp{P^\bullet}
is diffeomorphic to
$(C^{\bullet}\times \rea^2, ({\cal E}_5)_{C^{\bullet}\times\rea^2})$.
\end{lem}
Proof can be found in \cite[]{16af}. \cbdo

\end{appendix}
\newpage

%$$$$$$$$$$$$$$$$$$$$$$ REFERENCES $$$$$$$$$$$$$$$$$$$$$$$$

%\bibliographystyle{plainnat}
%\bibliography{artfiz1,books1}

\end{document}